# Who Pays for the KV Cache?

## Attributing Shared AI Inference Spend Across Kubernetes and LLM Provider Bills

Timothy Urista
Independent Researcher

**Abstract.** Organizations pay for AI through disconnected ledgers: Kubernetes allocations for self-hosted inference, gateway logs, and per-token bills from API providers. We present `unalloc`, an open-source tool that joins OpenCost, LiteLLM, OpenAI and Anthropic cost data into one Decimal-exact ledger and reports the share of spend with no owner, and use it to study where attribution breaks at the seams between these systems. Five case studies run inference for real or simulate it: a vLLM-style serving simulator with paged KV memory and prefix caching; a PyTorch transformer serving a multi-tenant trace with a real KV cache; tensor- and pipeline-parallel inference on `torch.distributed`; the unmodified CLI against mock provider APIs; and four downstream use cases. At the seams, in a constructed multi-pod deployment scenario — one month of synthetic OpenCost allocations, not observed organizational billing data — owner labels set only on Leader-WorkerSet leader pods leave **66%** of that deployment's GPU bill unowned, and the natural fallback key assigns **61%** of it to a Helm chart name while the headline unallocated share falls to 4%; enabling every source double counts all gateway spend; and reading one page of a billing API reports a quarter of spend. Inside a shared inference server the metering rule decides who pays: on an NVIDIA H100 running vLLM, a token meter assigns a retrieval-heavy tenant **12–14 percentage points** more of the bill than an equal time-share meter at every load tested, while GPU utilization reads 97–99% across configured loads of 2 to 16 requests per second (3.7 to 26.9 completed requests per second) and power draw tracks load. Neither meter is a ground truth, and we position these results against recent Shapley-based energy attribution. Code, raw data, captured evidence, figures and this paper regenerate from the repository.

## 1 Introduction

The cost of an AI feature is split across systems that were never designed to agree. A retrieval-augmented answer might touch a vector database and an embedding job on Kubernetes, a self-hosted open-weights model on a GPU pool, and a frontier model behind an API gateway. Tooling for this already exists and is improving: OpenCost [11] prices Kubernetes workloads, its plugin interface brings external provider costs — including an OpenAI plugin — alongside them, and its inference accounting work addresses allocation versus usage cost, token costs and cache effects. FOCUS [24] standardizes billing records across providers and represents split-cost allocation. What remains awkward in practice is the seam: the same dollar can arrive through more than one of these paths, ownership metadata is set per workload rather than per template, and a billing read that silently stops early looks identical to a cheap month. The question those seams raise is what share of the month's spend belongs to no one — the number FinOps allocation practice is built around [14].

`unalloc` exists to produce that number. Its design is deliberately small (§2): every source is parsed into one `CostRow` type, labels are canonicalized so `label_costCenter`, `team_id` and a `team:platform` request tag join, and attribution groups the joined ledger on a chosen owner key. The interesting questions are empirical. Shared inference servers batch requests from many tenants [2], share KV memory through paged allocation and prefix caching [1, 4], and scale across devices with tensor and pipeline parallelism [5, 6]. Each of these mechanisms moves cost between tenants, and each leaves a trace — or fails to — in the ledgers a tool like `unalloc` reads.

We ask four questions, each answered by a case study:

1. **Metering.** When one inference pod serves several tenants, how much does the choice of meter (tokens, list-price tokens, compute time, KV memory) change each tenant's bill? (§4, §5, §9)
2. **Distribution.** What happens to attribution when one model replica spans several pods? (§6)
3. **Joining.** What goes wrong when gateway and provider ledgers are combined? (§7)
4. **Use.** What can an organization do with a joined ledger beyond a percentage? (§8)

**Contributions.** (i) An open-source tool that joins Kubernetes, gateway and provider cost ledgers, with exact money handling, deterministic label canonicalization, invoice reconciliation, fallback-aware reporting and a CI budget gate. (ii) Attribution failures at the seams between those ledgers, measured end to end: label propagation in multi-pod serving, fallback misattribution, double counting and truncated billing reads (§6, §7). (iii) A comparison of metering rules for shared inference servers in simulation, with real PyTorch inference on CPU, and with vLLM on an H100 (§4, §5, §9), positioned against work that measures a Shapley reference for request energy (§10). (iv) A reproducible artifact: raw data, the captured GPU bring-up and teardown, and one-command regeneration. Defects the studies exposed in the tool itself, each now covered by a regression test, are listed in Appendix A.

## 2 The joined ledger

### 2.1 Model

A `CostRow` is one unit of spend: a source, a cost-object name, an amount in USD as a `Decimal`, a time window, normalized labels, and an optional quantity (tokens, GPU-hours). Adapters only parse; they never aggregate or decide attribution. Given rows $R$, an owner key $d$ and an ordered fallback list $f_1, ..., f_k$, attribution assigns each row the first non-empty value of $d, f_1, ..., f_k$, or `<unallocated>`:

$$U(R, d) = \frac{\sum_{r \in R,\ \mathrm{owner}(r)=\perp} \mathrm{amount}(r)}{\sum_{r \in R} \mathrm{amount}(r)}$$

Because fallbacks make the headline smaller without making it more correct, `unalloc` also reports $F$, the spend attributed **only** through a fallback key (§6 shows why this matters).

`CostRow` is deliberately not a billing interchange format. FOCUS [24] is the standard for that, with a far larger column set and an explicit representation of split-cost allocation; a FOCUS-conformant ledger would be a reasonable source for the analysis here. What we implement is the smaller application-specific shape that ownership analysis needs, and we claim no conformance: establishing it would require a field-level mapping we have not carried out.

### 2.2 Canonicalization

Label keys are lower-cased, camelCase is split, path-style keys keep their last segment (`app.kubernetes.io/name` → `name`), known provider prefixes are stripped until none remain, and an alias table maps spellings (`team_id`, `owner`, `squad`) onto one dimension. When distinct raw keys collapse onto one canonical key, a fixed precedence decides: a key already written canonically, then an alias, then a rewritten key, then a path-derived key, with ties broken on the raw key. Before the case studies the tool kept whichever key arrived first, which made results depend on provider serialization order.

### 2.3 Reconciliation and gates

`reconcile` compares each source's ledger total with a billed amount and flags deltas above a tolerance; `report --budget P` exits non-zero when $U > P$, so a deployment pipeline can refuse to ship unowned spend.

## 3 Methodology

Table 1 summarizes the studies. Every study emits payloads in the providers' own response shapes and parses them with `unalloc`'s production adapters; the end-to-end study goes further and runs the CLI binary against HTTP servers. Prices are deliberately round: one H100-class GPU at \$3.00/hour, OpenCost's default CPU and RAM rates, and illustrative per-million-token API prices. The questions are about **shares**, which do not depend on the absolute rates. Unless stated, a monthly pool is 8 GPUs × 720 h = \$17,280.

| Study | What runs | Real vs. modelled | Runtime |
|---|---|---|---|
| `kv_cache` | Discrete-event vLLM-style engine: paged KV blocks, prefix caching, chunked prefill, continuous batching, recompute preemption; 4 tenants, Poisson + multi-turn sessions | modelled latency | 102 s |
| `torch_kv` | 3.3M-parameter decoder (4 layers, $d = 256$, RoPE [9]) with explicit KV cache; 96-request 4-tenant trace on CPU | real inference | 29 s |
| `distributed` | Megatron-style TP (2, 4 ranks) and 2-stage PP on `torch.distributed`/gloo; LeaderWorkerSet cost scenarios | real inference | 24 s |
| `hybrid_e2e` | CLI subprocess vs. mock OpenCost, LiteLLM, OpenAI, Anthropic APIs with real auth schemes and cursor pagination | real CLI, synthetic spend | 2 s |
| `use_cases` | Labeling Pareto, feature unit economics, self-host break-even, CI gate | mixed | 25 s |
| `gpu_validation` | vLLM 0.29.0 serving Qwen2.5-7B-Instruct on one NVIDIA H100 80GB, driven with the `kv_cache` tenants at configured loads of 2–16 requests/s (§9) | real serving, synthetic traffic | 9 min |

Table 1: Case studies. Runtimes on an Apple-silicon laptop (10 cores), Python 3.11, PyTorch 2.14 CPU, except `gpu_validation`, which ran on a DigitalOcean GPU Droplet.

## 4 A shared vLLM pod: the meter decides who pays

### 4.1 Setup

The simulator models one replica of an 8-replica fleet serving an 8B-parameter model (131 KB of KV per token) from a 12,000-block pool (16 tokens per block). Engine steps share a 2,048-token budget between decodes and chunked prefills [3], and step latency is an analytic function of prefill tokens, decode sequences and attended context. Four tenants arrive by Poisson process: **search** (RAG: 600–1,400-token prompts, 16–64-token answers, three shared system prompts), **agents** (multi-turn sessions that re-send the whole conversation each turn, 150–400-token outputs), **platform** (medium/medium) and an unlabeled shared **sandbox** key. Each step's duration is split across the sequences in it by their marginal work; each KV block-second is split across the requests holding the block. Idle time and unheld blocks are recorded separately as overhead.

At 3 requests/s for 30 minutes the engine served 10,020 requests with a 75% prefix-cache hit rate, no preemptions, and a median time-to-first-token of 38–96 ms per tenant. (At 4 requests/s a five-minute run looks healthy but agent follow-up turns accumulate over half an hour and median TTFT exceeds 40 s; we report the sustainable operating point.)

### 4.2 Results

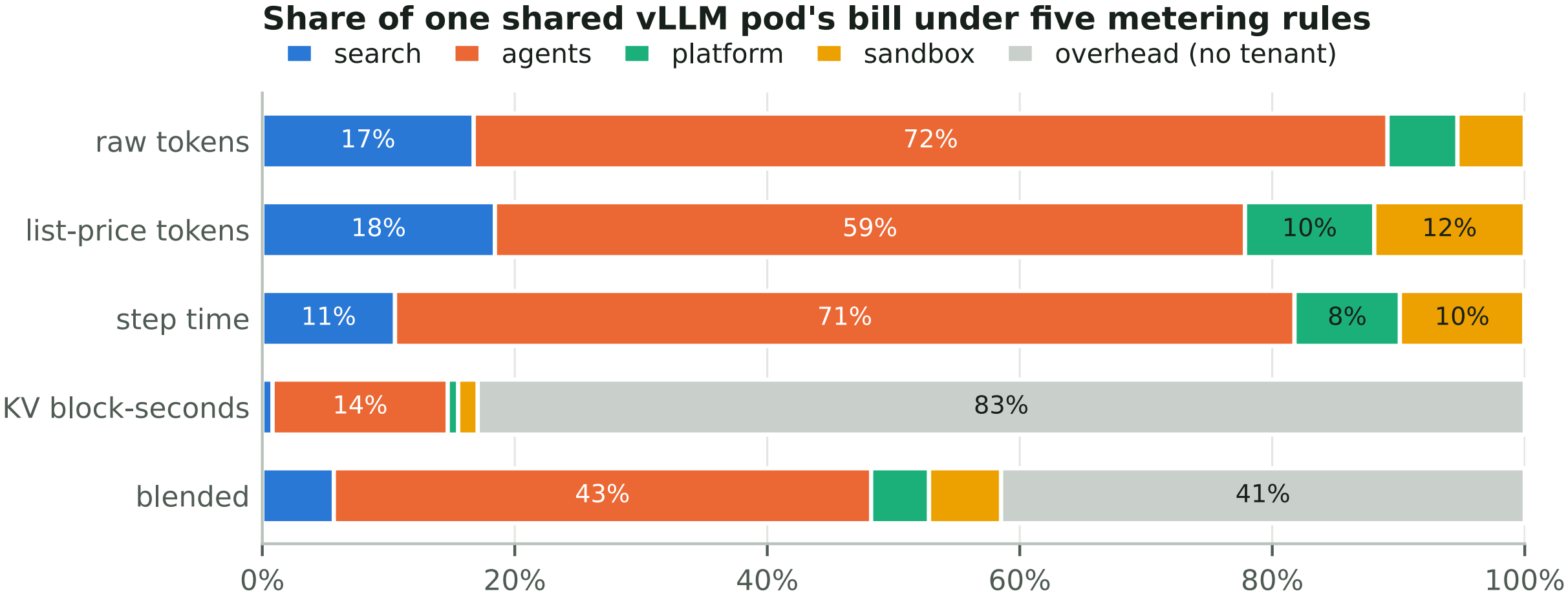

Figure 1: One pod bill, five meters. Token meters have no notion of overhead. Step-time metering sees none either once continuous batching keeps any request in flight; KV-memory metering leaves 83% of the bill with no request.

Figure 1 shows the split. Agents dominate every compute-like meter because each turn re-sends a growing conversation, but the meters still disagree: raw tokens charge **search** 16.7% of the pod, step time 10.5% — a 6.2-point gap that grows to 12.0 points against KV memory once overhead is redistributed. List-price weighting (cached input at 0.1×, output at 4×) is no closer: it moves agents from 72.4% to 59.4%, twelve points away from measured step time (71.2%).

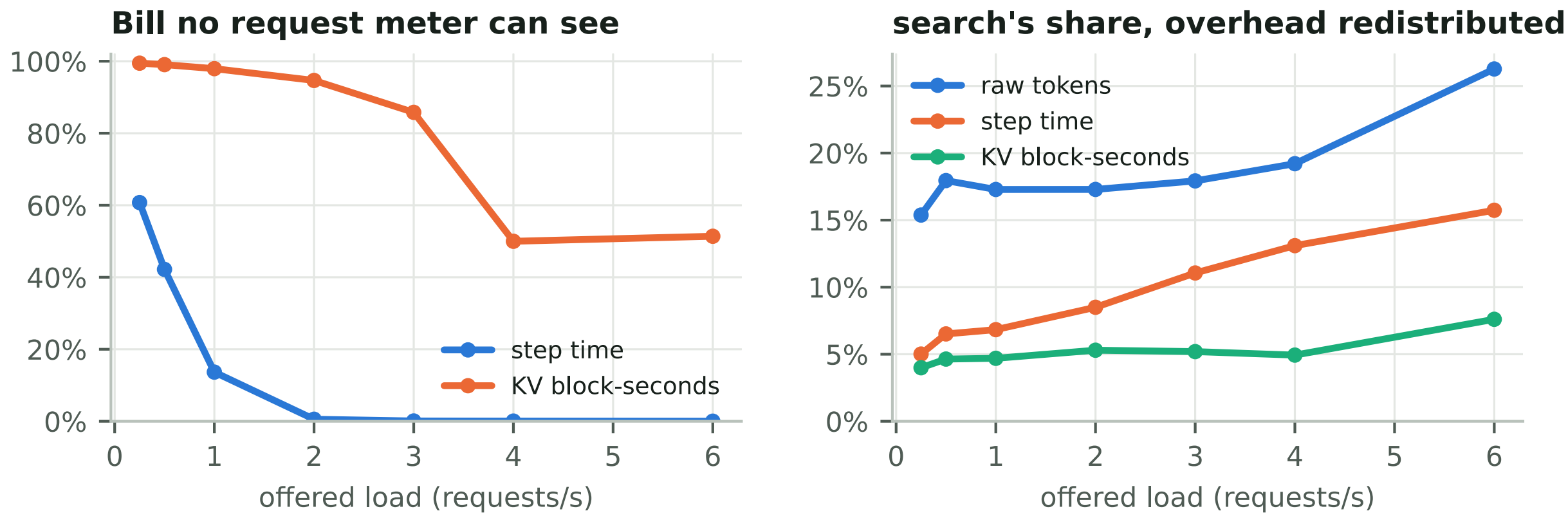

Figure 2: Left: overhead visible to each meter as load rises. Right: search's share under three meters, overhead redistributed.

Figure 2 explains why. Step-time overhead collapses to ≈0 by 2 requests/s: a pod that is "busy" in wall-clock terms is nowhere near its token capacity (754 vs. ≈1,340 output tokens/s). KV memory tells the opposite story — 95% of block-seconds are unheld at the same load, because pools are provisioned for peaks. Neither is wrong; they measure different scarce resources, and the choice moves several points of the bill between tenants. §9 checks both observations against a real vLLM server on an H100.

> **For `unalloc`.** A per-tenant split only closes the gap if overhead is labeled. Emitting the memory split with an unlabeled overhead row reports 84.5% unallocated; redistributing the same overhead as OpenCost `sharedCost` reports 9.1% — only the sandbox key. The unallocated percentage is as much a statement about the overhead policy as about labels.

## 5 Real inference with a KV cache

### 5.1 Setup

A from-scratch decoder-only transformer (3.28M parameters, random weights with a fixed seed — serving cost does not depend on weight values) implements a preallocated per-layer KV cache and shared-prefix reuse. Before any timing, cached decoding is checked against full recomputation: generated tokens are identical and the maximum logit difference is $1.07 \times 10^{-6}$. The trace has 96 requests from the same four tenant shapes; it is served three times and per-request medians are used. Measured shares moved by at most 4.3 points across repeats.

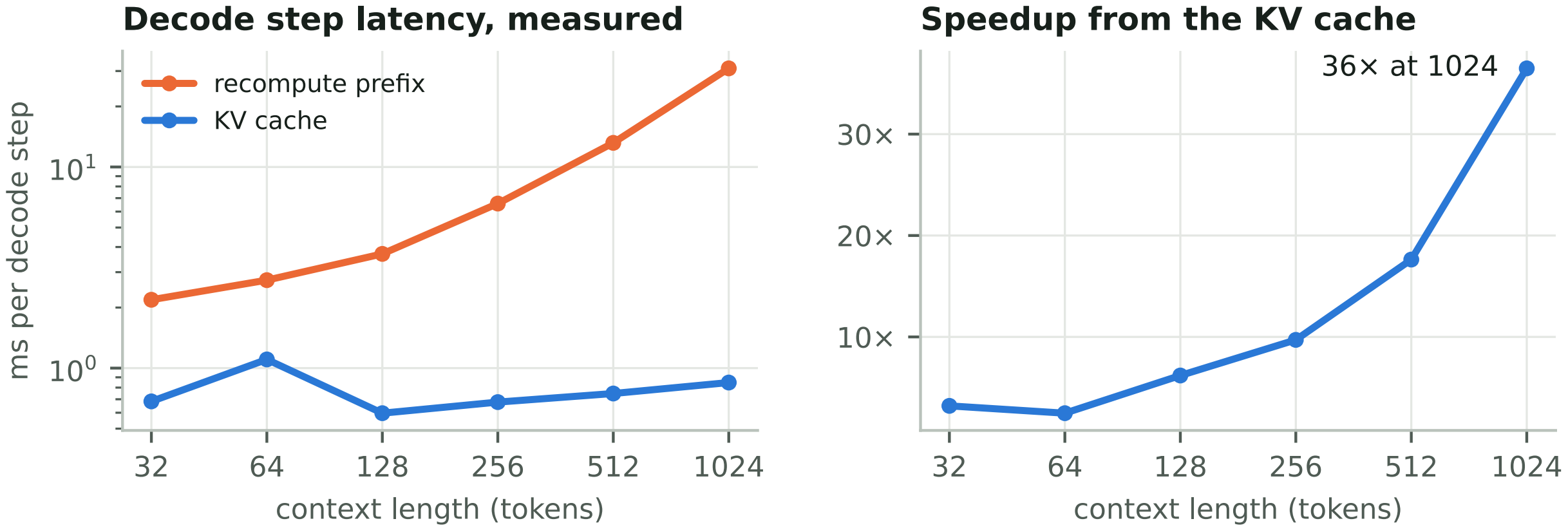

Figure 3: Measured decode-step latency on CPU. At 1,024 tokens of context the cache is 8.0 MiB and a step takes 0.85 ms instead of 30.9 ms.

### 5.2 Results

Figure 3 shows the expected shape [7]: recomputation grows with context while a cached step avoids it, 36.5× faster at 1,024 tokens. Over the 32–1,024-token range measured here, cached-step latency changes far less than full recomputation; it should not be read as independent of context length in general, and Pope et al. explicitly discuss generation latency rising as the KV cache grows. The attribution consequences are in Figure 4. Search sent 18,024 prompt tokens and received 701; agents sent 2,345 and received 4,017. Token counting (and analytic FLOPs [8], which track tokens) charge search ≈45% of the pool. Measured compute charges it 12.4%: prefill is batched and cheap, decode is sequential and expensive.

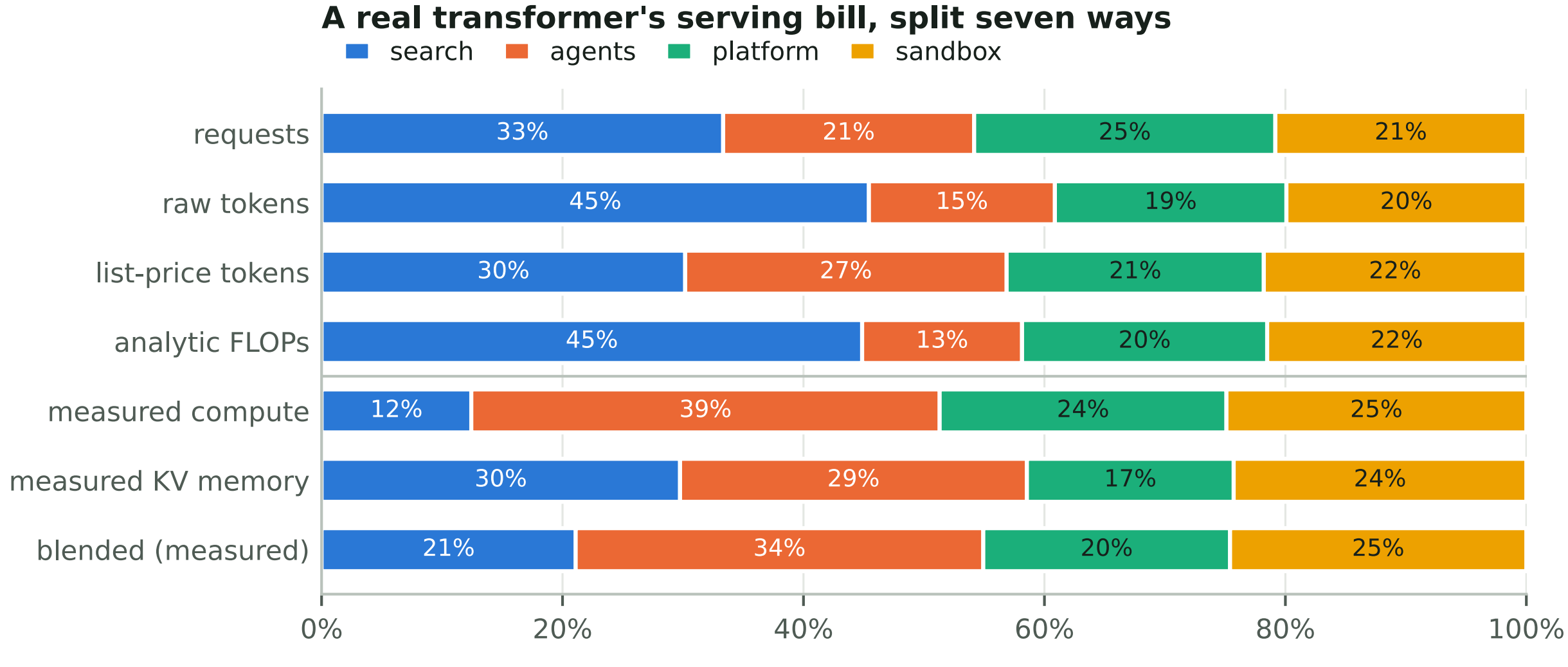

Figure 4: The same measured trace split seven ways. Above the rule: meters derivable from request logs. Below: meters that require serving telemetry.

**Headline.** Token counting and measured compute disagree on the RAG tenant's share by **33.0 points** of the pool, $5,703 of $17,280 a month. Measured cost per token differs 9.2× between search and agents, which a flat token price charges identically; weighting output at 4× halves the disagreement (17.8 points) without closing it. Neither meter is a ground truth, and on a batching GPU server the disagreement is smaller (§9).

On a GPU the per-step fixed overhead is amortized across a batch, so the magnitude is hardware-specific. In these workloads token-based allocation assigns the prompt-heavy tenant a larger share than the timing-based meters. That phase asymmetry is consistent with the motivation for disaggregated serving [10], which separates prefill from decode because their resource profiles differ; the size and direction of the allocation difference, however, depend on the workload, the batching behaviour and the reference meter chosen, and we do not claim a universal subsidy from prompt-heavy to decode-heavy tenants.

# 6 Distributed inference: attribution leaks through pod templates

## 6.1 Parallel inference, verified

We shard the reference model Megatron-style [5] — column-parallel QKV and MLP up-projection, row-parallel output and down-projection with an all-reduce after each — and separately split it into two pipeline stages that pass activations with send/recv [6]. All configurations generate the same tokens as the single-process model (maximum logit difference $3.3 \times 10^{-6}$ for TP, exactly 0 for PP). Per-rank parameters drop from 13.65 MB to 7.36 MB (TP2) and 4.21 MB (TP4); embeddings, norms and the LM head stay replicated. KV cache per rank halves with each doubling.

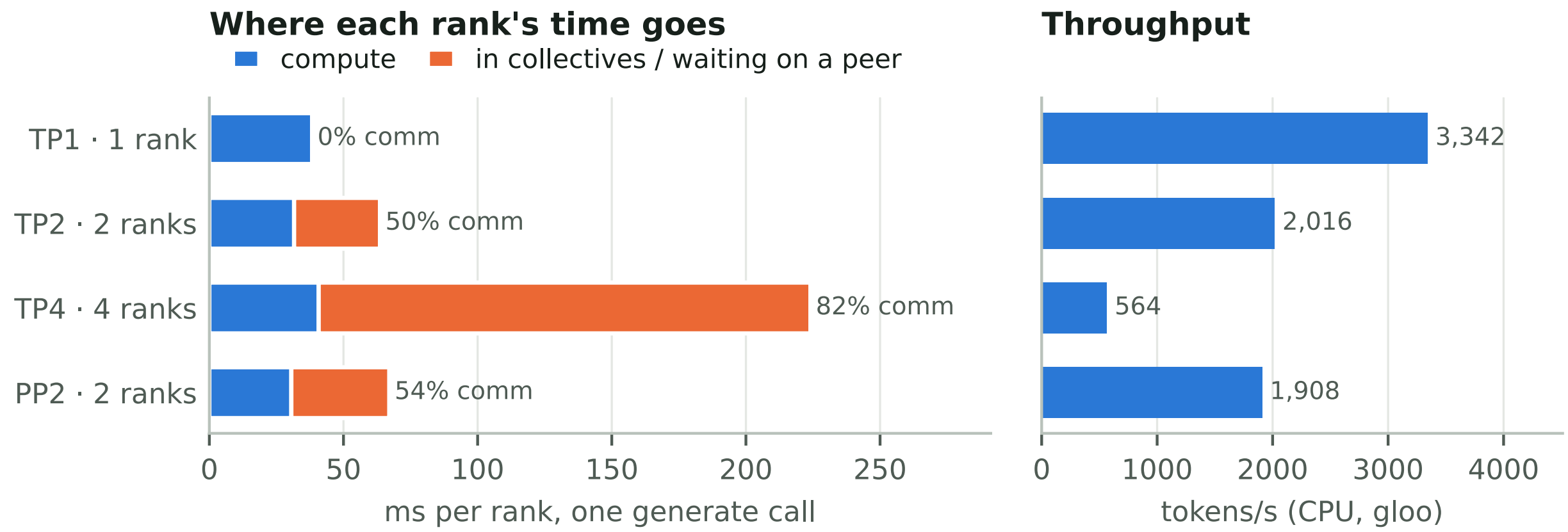

Figure 5: Per-rank time split for batch 4, 64-token prompt, 32 new tokens. Communication is time blocked in collectives or waiting for a peer.

On CPU over loopback, distribution is slower, as expected (Figure 5): TP4 spends 82% of each rank's time in collectives. These are CPU-over-loopback measurements; we have not measured the equivalent fractions on NVLink-connected GPUs, and we make no claim that they bound them. We use them only to size the scenario below, and report a sensitivity series at 5–20% because the attribution result should not depend on the fraction we picked.

## 6.2 The LeaderWorkerSet month

Multi-host serving on Kubernetes commonly runs one replica as a group of pods, e.g. with LeaderWorkerSet [13]: a leader template and a worker template. The failure below is conditional, not inevitable — if the leader template is omitted entirely the worker template applies to the leader too, and labels on it reach every pod. It arises when both templates are specified and only one carries the owner label, which is the shape a chart that sets resources or affinities separately per role tends to produce. We emit one month of OpenCost allocations for two tenants — **search** (2 replicas × 4 pods, tensor parallel) and **agents** (3 replicas × 2 pods, pipeline parallel) — plus cluster idle, $38,400 in total, with each pod's GPU time split into compute, communication and in-pod idle using the measured fractions.

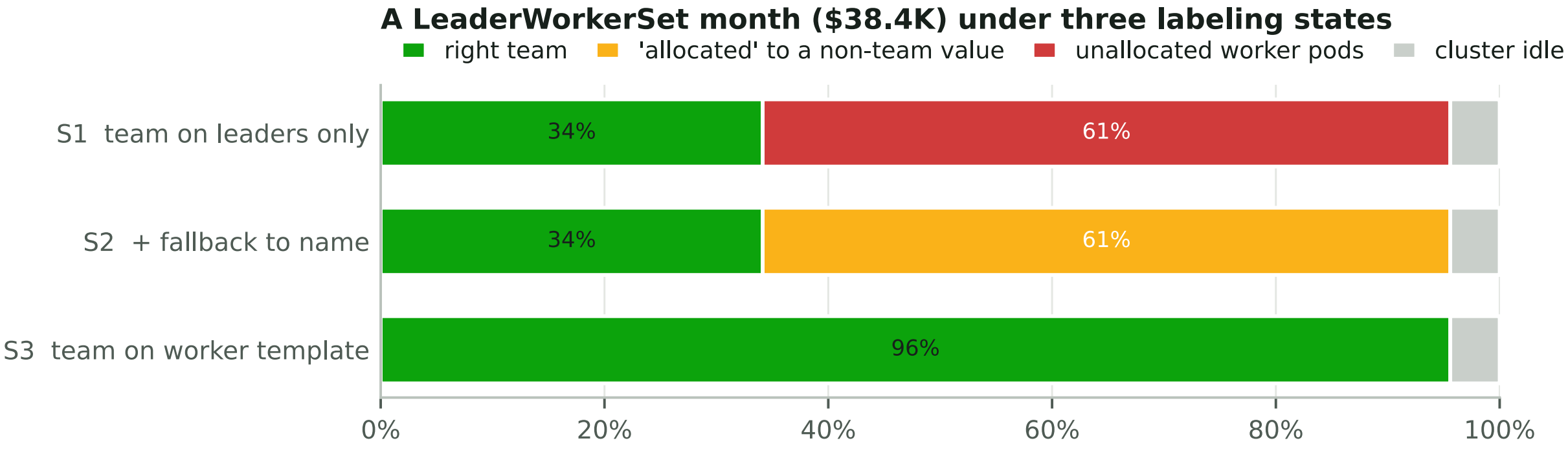

Figure 6: Correctness of attribution, not just coverage, under three labeling states.

In S1 the `team` label is on the leader template only: 65.9% of the month is unallocated. The obvious rescue is a fallback key every pod carries. In S2, falling back to `name` drops the unallocated share to 4.4% — and routes $23,597 (61%) into a bucket called `vllm`, the Helm chart's `app.kubernetes.io/name`, which collides with `leaderworkerset.sigs.k8s.io/name` on the canonical key `name`. S3 fixes the worker template: 95.6% correct, only idle remains.

> **For `unalloc`.** The headline percentage cannot distinguish S2 from S3. We added the fallback-attributed amount $F$ to every report ($23,597 in S2, $0 in S3) and made collision resolution order-independent; before the fix the S2 bucket was `vllm` or `search-llama-70b` depending on JSON key order.

Communication is itself a cost with an owner. GPU time spent in collectives totals $12,059, $8,201 of it on worker pods that S1 leaves unowned. A per-token showback, which spreads it by token share, moves $3,737 from the tensor-parallel tenant onto the pipeline-parallel one; at a more GPU-like 10% communication fraction the transfer is $166.

## 7 Joining gateway and provider ledgers

The organization in this study runs a LiteLLM gateway [12] in front of OpenAI and Anthropic, the self-hosted fleet above, and — like most — a research team calling providers directly. Mock servers enforce each provider's authentication (Bearer for OpenAI and LiteLLM, `x-api-key` for Anthropic's Admin API) and page responses with cursors; in the dev container the LiteLLM spend log lives in Postgres, as it does in production. The CLI runs as a subprocess with only environment variables pointing it at the servers.

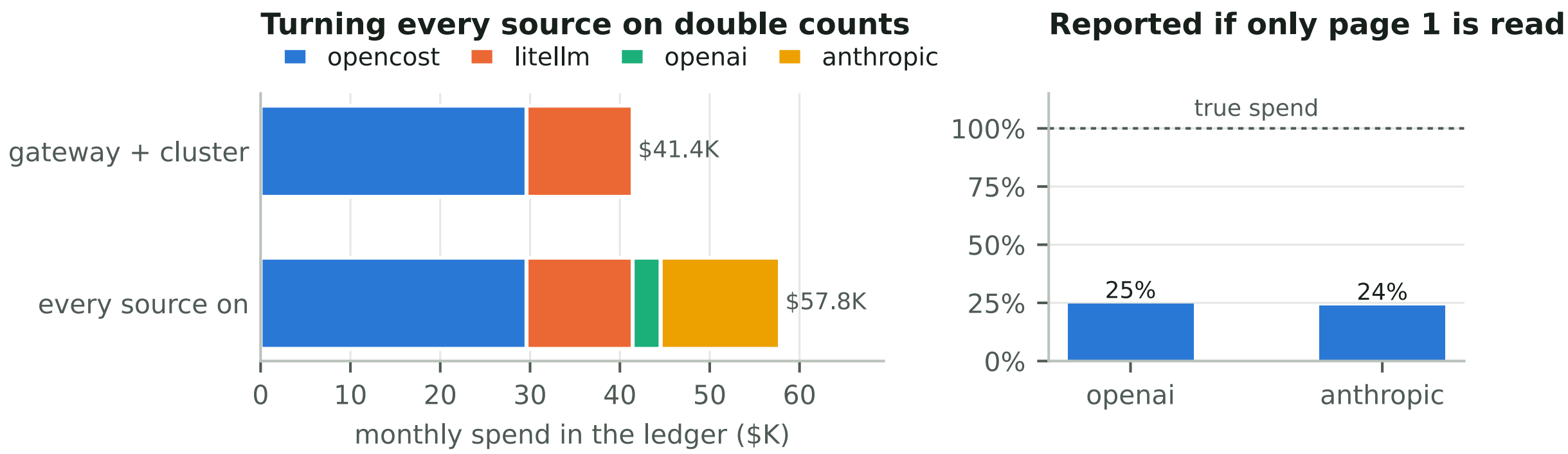

Figure 7: Left: ledger totals for the recommended sources versus every source. Right: share of true provider spend reported by an adapter that reads only the first page.

| Scenario | Ledger total | Unallocated |
|---|---|---|
| A. Cluster + gateway (`--source opencost --source litellm`) | $41,420 | 66.9% |
| B. Every source enabled | $57,813 | 76.3% |
| C. Provider billing only | $16,393 | 100.0% |
| C′. Provider billing, fallback to project/workspace | $16,393 | 0.0% |
| Provider ledgers vs. provider invoices | matched | 0.0% delta |
| Gateway ledger vs. provider invoices | $11,815 of $16,393 | −27.9% |

Table 2: End-to-end scenarios over one month.

Table 2 and Figure 7 give three results. **Double counting:** enabling every source adds $16,393, of which $11,815 — exactly the gateway's spend — is counted twice; the remainder is real bypass traffic the gateway never saw. **Coverage:** the gateway ledger reconciles to only 72.1% of the provider invoices, so a gateway-only view under-reports by $4,578 and cannot know it without reconciliation. **Fallback illusion:** provider billing alone has no team dimension; falling back to project and workspace reports 0% unallocated while attributing nothing to a team.

Before the fixes in Table 4, this study could not run at all: the OpenAI and Anthropic adapters ignored their default URLs whenever the CLI passed an unset environment variable, Anthropic rejected the Bearer header, and neither adapter followed pagination — reading 25% of OpenAI spend and 24% of Anthropic spend.

## 8 Use cases for a joined ledger

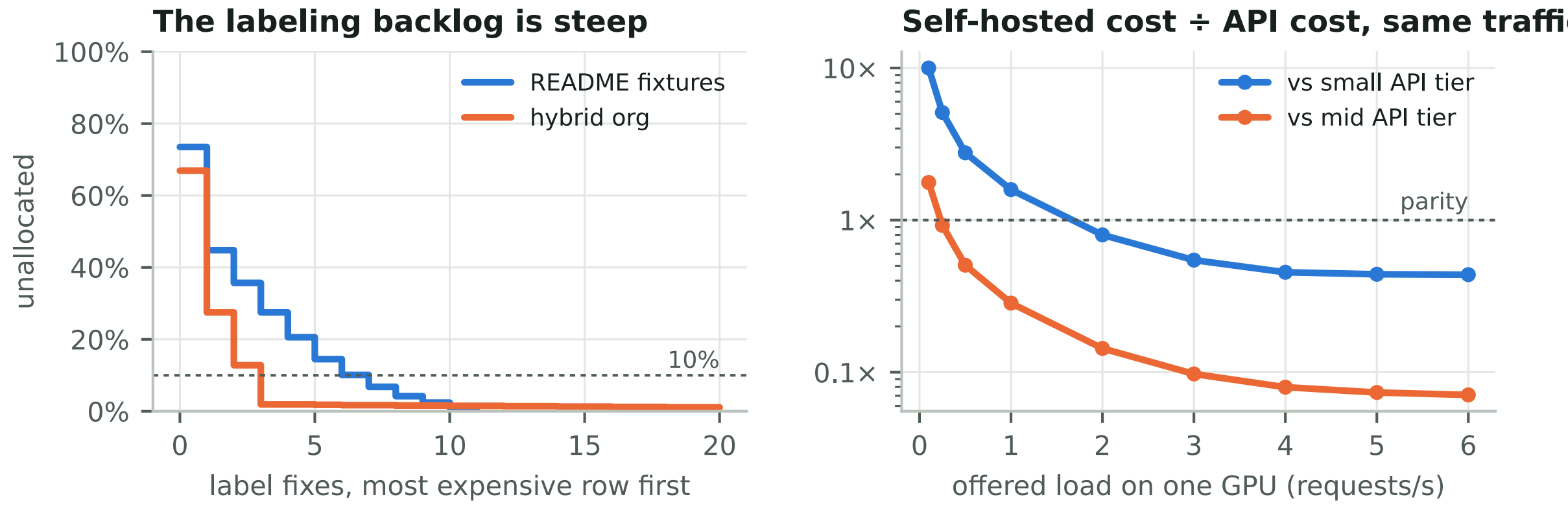

Figure 8: Left: unallocated share after fixing the most expensive unlabeled rows first. Right: cost of serving the simulated traffic on one self-hosted GPU divided by the same tokens at API list prices, for two illustrative price tiers.

**Labeling is a Pareto problem.** Ordering the backlog by dollars (Figure 8, left), three label changes take the hybrid organization from 66.9% to 1.9% unallocated; the README fixtures need seven changes to cross 10%. A dollar-ordered backlog turns a governance program into a short ticket list.

**Feature unit economics need both ledgers.** Joining on a `feature` label, the RAG **answer** feature costs $22.32 per thousand requests including its vector database, against $13.38 if only the LLM bill is counted: 40% of the feature's cost is invisible to a gateway-only view.

**Build-versus-buy depends on utilization, not list price.** One self-hosted GPU serving the simulated traffic is cheaper than a mid-tier API above ≈0.23 requests/s and cheaper than a small-tier API above ≈1.7 requests/s (Figure 8, right), flattening at 0.08× and 0.45× respectively once saturated. The comparison bounds cost, not quality.

**Budgets belong in CI.** `unalloc report --budget 50` exits 2 on the fixtures (73.5% unallocated); `--budget 80` exits 0, so a deployment that ships unlabeled workloads can fail review.

# 9 Validation on a datacenter GPU

## 9.1 Setup

The CPU timings of §5 and the analytic latency of §4 leave one question open: does the metering result survive production serving software on production hardware? One DigitalOcean GPU Droplet (`gpu-h100x1-80gb`: NVIDIA H100 80GB HBM3, driver 580.173.02, CUDA 13.0, 20-vCPU Xeon Platinum 8468, 235 GB RAM) ran vLLM 0.29.0 [1] serving Qwen2.5-7B-Instruct in bf16 with an 8,192-token context and prefix caching; vLLM sized its KV cache at 995,296 tokens. The model is Qwen2.5-7B-Instruct as published on Hugging Face [25]. vLLM 0.29.0 is a release [26] far newer than the PagedAttention paper [1] that introduced the engine, and the captured server log in the repository records the build actually served.

A client on the same machine replayed the §4 tenants as token-id prompts, so lengths were exact and shared system prompts byte-identical, with output lengths forced and greedy decoding. It ran two minutes each at configured loads of 2, 4, 8 and 16 requests/s. The configured rate is the Poisson arrival rate of **session-initial** requests; the follow-up turns of multi-turn sessions arrive after their predecessor completes and are not counted in it, so total completed traffic was higher — 3.7, 7.1, 13.4 and 26.9 requests/s respectively (Table 3). We report the configured rate as the load setting and completed throughput separately; every per-request and per-token result below is computed over all completed requests. The client recorded each request's time to first token, completion time and the server-reported cached-token count, and sampling vLLM's Prometheus metrics twice a second and `nvidia-smi` once a second. The time-share meter splits every 50 ms of wall time equally across in-flight requests. The droplet existed for 29 minutes (about $2.15); the runbook and the captured terminal output of every step, from creation to a verified deletion, are in the repository.

## 9.2 Results

| **Load** req/s | **Requests** (done/s) | **Output** tok/s | **Cache** hits | **TTFT** p50 / p95 ms | **TPOT** p50 ms | **GPU** util | **Power** W | **search** tokens | **search** time |
|---|---|---|---|---|---|---|---|---|---|
| 2 | 446 (3.7) | 773 | 67% | 28 / 44 | 6.3 | 97% | 469 | 16.5% | 4.8% |
| 4 | 862 (7.1) | 1,479 | 79% | 26 / 43 | 6.7 | 99% | 499 | 16.6% | 4.7% |
| 8 | 1,638 (13.4) | 2,761 | 78% | 29 / 47 | 7.4 | 99% | 547 | 17.2% | 4.7% |
| 16 | 3,295 (26.9) | 5,389 | 66% | 51 / 99 | 12.2 | 99% | 660 | 18.9% | 5.3% |

Table 3: vLLM on an H100. **Load** is the configured arrival rate of session-initial requests; with follow-up turns the completed rate in parentheses is roughly double it. All 6,241 requests completed without error, and no queued request was observed in any half-second telemetry sample. TTFT is time to first token; TPOT is time per output token. The last two columns are search's share of the bill under token and time-share meters, overhead redistributed.

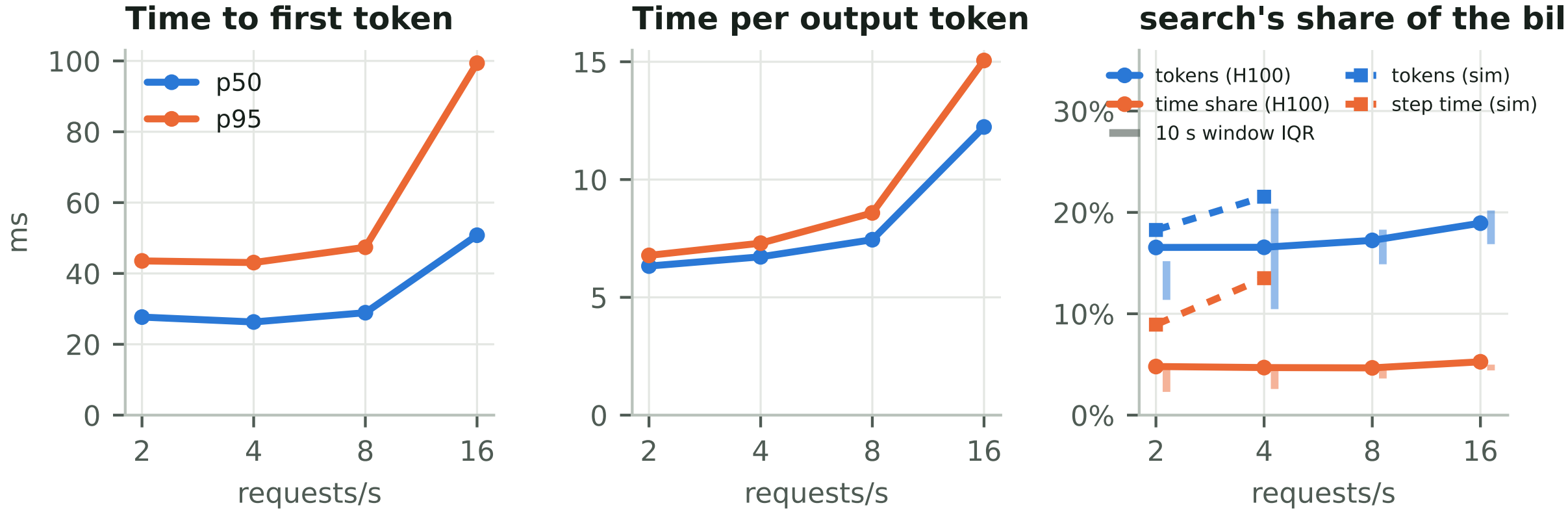

Figure 9: Left, middle: latency percentiles by load on the H100. Right: search's share of the bill under token and time-share meters on the GPU (solid), and under the simulator's token and step-time meters where the simulator is not saturated (dashed). The vertical bar beside each H100 point is the interquartile range of that meter recomputed in ten-second windows of the same run — a description of within-run traffic variation, not an uncertainty estimate. It is drawn separately because the whole-run share is traffic-weighted and need not lie inside the quartiles of its windows.

**The meters still disagree.** The token meter assigns search 16.5–18.9% of the bill; the time-share meter assigns 4.7–5.3%, a disagreement of 11.7–13.7 points at every load (Table 3, Figure 9). At the 2 requests/s load the simulator predicted 18.3% against 8.9%, a 9.4-point disagreement. The CPU experiment of §5 measured a larger gap, 33 points, on the same tenant mix. That is not a controlled comparison — the model, the serving software, the hardware and the execution of the workload all differ — so we report it as a difference between two experiments rather than an effect with a measured cause. Our hypothesis is that an unbatched decode step on CPU carries large fixed overhead that a batching GPU server amortizes, which would compress the gap without changing its direction; testing it would need the same model and workload run both ways. Equal time share is itself a heuristic: it charges a request waiting on prefill the same as one decoding. Replaying tenant subsets, as JouleShare does for request energy [15], could provide a measured Shapley reference [19] for a specified cost function such as active GPU energy; with four tenants that is 15 non-empty coalitions per load level, before repetitions and baseline measurements. Translating such a reference into a division of a fixed rental bill would still require an overhead-allocation policy, so it bounds the metering question rather than closing it.

**How steady is that gap within a run?** One two-minute run per load level carries no error bar, so we recomputed both meters inside consecutive ten-second windows of each run, normalizing within the window and discarding the first one: while the request pipeline fills, the requests that have **completed** are disproportionately the short ones, which biases any share computed by completion time, and that window sits 3–4× above the rest at every load. Over the eleven remaining windows the token-versus-time divergence has a median of 9.3, 11.7, 12.0 and 12.8 points at the four loads, with an interquartile range of 2.5 to 7.5 points, and every window at every load is positive. The low-load numbers move the most, because a ten-second window there holds only about 35 completed requests.

These are descriptive statistics of one run, not an estimate of uncertainty. The windows are consecutive slices of a single traffic sample rather than independent observations, so we do not attach a confidence interval to them, and nothing here bounds how far a second run at the same load would land. Figure 9 draws the same quartiles beside each meter, separately from the whole-run point: the whole-run share is weighted by each window's traffic, so it need not fall inside the quartiles of its own windows, and at three of the eight points it does not. Run-to-run and seed-to-seed variation needs repeated runs, which the harness now supports (§11) and this dataset does not contain.

**Utilization is not a cost signal.** `nvidia-smi` reported 97% utilization at the 2 requests/s load and 99% at 4, 8 and 16, while throughput rose 7×. vLLM had at least one running request in 97–100% of samples at every load, and the KV cache was 0.7–8.1% occupied. This is the §4 overhead result observed directly:

time- and utilization-based meters see no idle capacity, and memory-based meters see almost nothing else. Power draw tracked load, from 469 W to 660 W.

**The simulator's latency was too pessimistic.** Below saturation its throughput matched (737 vs. 773 output tokens/s at the 2 requests/s load), but its step-latency constants saturate it by the 8 requests/s load, with multi-second time to first token, while the H100 served the 16 requests/s load — 26.9 completed requests/s — with a 99 ms p95. The served model also stores about 57 KB of KV per token against the simulator's 131 KB. The §4 shares should be read as directional and its latencies as uncalibrated; fitting the simulator's constants to this run is future work.

> **Operational.** vLLM 0.29.0′s default FlashInfer sampler failed to JIT-compile on the provider's GPU image and stopped the engine; the PyTorch sampler, equivalent for greedy decoding, worked. Deleting the droplet by tag cleared the tag before the droplet itself was gone, so teardown was verified by resource ID, not tag.

**Further use cases.** The same ledger supports multi-LoRA serving (attribution by adapter label on a shared base model), fine-tuning and batch embedding jobs whose GPU hours share clusters with online inference, agent systems whose tool calls fan out across providers, speculative decoding where a draft model's cost belongs to the target model's callers, and multi-cloud estates where Bedrock and Azure OpenAI bills join the same dimension.

## 10 Related work

**Attributing shared inference.** JouleShare [15] is the closest work to §4, §5 and §9. It measures exact request-level Shapley energy [19] on vLLM by replaying every subset of eight requests, finds that token-proportional attribution misallocates roughly a quarter of batch energy under both static and continuous batching, and fits a lightweight estimator to the measured Shapley shares. Its reference is exact for the quantity it defines — active GPU energy, with idle power subtracted, divided by an explicitly chosen Shapley rule — and that is a stronger footing than this paper has. It is not, and does not claim to be, a uniquely correct division of a fixed rental bill, which must also place idle capacity. We compare metering rules without any such reference, at the level of tenants rather than requests, and include KV-memory, list-price and time-share rules alongside tokens. Vellaisamy et al. [16] decompose inference energy on H100 and H200 GPUs into fixed and per-token components and show that lower energy per token can reflect amortization rather than lower total energy: the same mismatch between token pricing and resource consumption that our meters express in dollars. LLMVisor [17] decomposes co-batched latency into per-request contributions for fractional sharing of a serving engine, again outperforming token-count baselines. PrefixShield [18] makes tenant groups responsible for the prefix-cache blocks they materialize, a scheduling answer to the shared-KV question our memory meter only measures; its prototype isolates prefix lookup by accounting group, leaving cross-group reuse — and how to divide the cost of a prefix two tenants both benefit from — outside its scope. Shapley values [19] and dominant resource fairness [20] are the classical foundations for sharing joint costs and multi-resource capacity.

**Cost and emissions allocation in clouds.** Schneider and Mattia [21] allocate the energy and emissions of shared data-center machines, infrastructure and software to cloud users, combining resource reservations with hourly usage rather than machine measurements alone. ABACUS [22] proposes a FinOps service for budget monitoring and enforcement and discusses integration with infrastructure-as-code cost checks; its paper presents policy-engine integration as an extension and predictive methods as future work. `unalloc report --budget` is a narrower gate on a different quantity: the fraction of observed spend that carries no ownership metadata, rather than an absolute spending limit. Cost-Governed RAG [23] attributes per-tenant cost across embedding, retrieval and generation, integrating a vector store and an LLM gateway inside a common governance boundary, with vector-index memory linear per tenant. Our feature economics (§8) join the same layers across sources that share neither a governance boundary nor a key: Kubernetes allocations [11], a gateway [12] and provider billing. FinOps allocation practice [14] supplies the organizational target all of these serve.

**Cost tooling and standards.** OpenCost [11] supports Kubernetes allocation, external-cost plugins including one for OpenAI, and inference-specific accounting that discusses allocation versus usage cost, input and output token cost and cache effects. FOCUS [24] standardizes billing data across providers and includes representations of split-cost allocation. Our internal `CostRow` is a smaller, application-specific representation built for ownership analysis; this work does not establish FOCUS conformance, which would require a field-level mapping we have not done.

**What this paper adds.** Serving systems [1, 2, 3, 4, 10] define the mechanisms whose cost is attributed here but do not address billing, and the attribution work above measures cost inside one system. We build on the cost-tooling ecosystem rather than replacing it: a lightweight ledger spanning OpenCost, LiteLLM and direct provider billing, and a study of how source overlap, incomplete reads and ownership metadata change the answer. The contribution is the implementation and the reproducible characterization of these failure modes in this workflow — label propagation across multi-pod serving templates, fallback keys that resolve to non-owners, double counting and truncated billing reads — not the general idea of consolidating cost data, and not a claim that no other tool reads the same four sources.

## 11 Discussion and limitations

**Recommendations.** (1) Decide the meter explicitly and publish it; token counts are a pricing choice, not a measurement. (2) Keep overhead — idle GPUs, empty KV pools, communication — as its own labeled line and redistribute it by policy, rather than letting any meter absorb it silently. (3) Put owner labels on every pod template of a multi-pod workload, and treat fallback-attributed spend as unverified. (4) Choose sources so each dollar has one path into the ledger, and reconcile against invoices every month.

**Limitations.** No allocation in this paper is a ground truth: the meters compared are defensible rules, and what we measure is how far they disagree, not which share is correct. The PyTorch timings of §5 and §6 are from CPU, where fixed per-step overhead and loopback collectives inflate decode and communication costs. §9 re-measures the metering result with vLLM on an H100, where its direction holds and its size is smaller. That run covers one GPU, one model, one two-minute run per load level and synthetic traffic, and its multi-turn prompts append synthetic assistant tokens, so cross-turn prefix reuse excludes previous answers. The ten-second window analysis in §9 describes the variation within each run and is not an uncertainty estimate: its windows are consecutive slices of one traffic sample, not independent observations. It cannot see run-to-run variance, warm-up effects that persist across a whole run, or anything specific to this droplet, this driver or this model; a campaign of repeated runs per load level — which `bench.py --repeats` now performs, interleaving the load levels so drift during the session does not land on one of them — is the experiment that would bound those, and we have not yet paid for it. Treat the 12–14 point result as one well-instrumented observation, not an estimate with a known standard error. The serving simulator uses analytic step latency whose constants proved too pessimistic for that GPU. Dollar figures use round, illustrative prices. The mock provider APIs reproduce authentication and pagination semantics as documented, not every field of the live responses.

## 12 Reproducibility

Everything in this paper regenerates from the repository, archived as release 0.2.3 at doi:10.5281/zenodo.22761012:

```
make research          # CPU PyTorch, notebook tooling, typst
make case-studies      # writes case_studies/results/*/metrics.json
make paper             # figures + this PDF
make notebook          # executes the companion notebook
make devcontainer-check  # the same, inside the isolated dev container
python -m case_studies.gpu_validation.analyze   # §9, from the committed raw GPU data
```

The GPU run itself follows `case_studies/gpu_validation/RUNBOOK.md` on any single-GPU machine, after a free dry run against a bundled fake server. `bench.py --repeats n` runs each load level $n$ times with a fresh seed, interleaving the load levels, and the analysis reports the spread across those runs alongside the within-run window statistics; the dataset published here is $n = 1$, which is why §11 treats the result as an observation rather than an estimate. A browser-based ledger explorer (`python -m case_studies.ui`) shows every dataset row by row with the owner each row resolves to, and a companion notebook walks through each study.

## Use of AI tools

The case-study code, experiment orchestration, analysis scripts, figures and drafts of this text were produced with Claude Code (Anthropic, Claude Opus 5) working under the author's direction, including the GPU run in §9. Generative AI is not an author of this work; the author is responsible for all of its content.

## A Defects found by the case studies

The case studies exposed these defects in `unalloc` itself. Each is fixed and covered by a regression test; they are listed here rather than as findings because they concern the tool, not the attribution questions the paper studies.

| **Area** | **Defect** | **Found by** |
|---|---|---|
| Fetch | OpenAI/Anthropic default base URL overridden by an unset env var; every live fetch failed | review; e2e |
| Fetch | Anthropic Admin API called with Bearer auth instead of `x-api-key` + `anthropic-version` | review; e2e |
| Fetch | No cursor pagination on provider cost APIs; first page only (≈25% of spend) | review; e2e |
| Money | OpenCost component costs summed through `float` when `totalCost` was absent | review |
| Labels | Only one provider prefix stripped; `label_app_kubernetes_io_name` ≠ `app.kubernetes.io/name` | distributed |
| Labels | Canonical-key collisions resolved by arrival order, incl. explicit key vs. alias | distributed, e2e |
| Labels | OpenCost annotations silently overrode same-key labels | distributed |
| Report | `labels` backlog ignored `--fallback`, disagreeing with `report` | review |
| Report | `labels` printed “Nothing unallocated” when no source loaded | review |

Table 4: Defects fixed in the course of the studies. Each has a regression test.